\documentclass[%
aps,
jmp,%
amsmath,amssymb,
reprint,%
]{revtex4-1}
\usepackage{graphicx}
\usepackage{multirow}

\usepackage{diagbox}
\usepackage{subfigure}
\usepackage{dcolumn}
\usepackage{bm}
\usepackage{epstopdf}
\usepackage{amsmath}
\usepackage[colorlinks]{hyperref}
\hypersetup{
	colorlinks = true,
	citecolor = {blue},
	linkcolor = {red}
}

\newcommand{\blk}{\color{black}}

\begin{document}
	\title{Qubit-based distributed frame synchronization for quantum key distribution}
	\author{ Ye Chen,$^{1}$ Chunfeng Huang$^{1}$, Shuyi Huang$^{1}$,  Zhenrong Zhang$^{2}$ and Kejin Wei$^{1,*}$ }
	
	\address{
		$^1$Guangxi Key Laboratory for Relativistic Astrophysics, School of Physical Science and Technology,
		Guangxi University, Nanning 530004, China\\		
			$^2$Guangxi Key Laboratory of Multimedia Communications and Network Technology, School of Computer Electronics and Information, Guangxi University, Nanning 530004, China\\
		$^*$kjwei@gxu.edu.cn
	}
	\date{\today}
	
	\begin{abstract}
		
			Quantum key distribution (QKD) is a method that enables two remote parties to share a secure key string. Clock synchronization between two parties is a crucial step in the normal operation of QKD. Qubit-based synchronization can achieve clock synchronization by transmitting quantum states between two remote parties, eliminating the necessity for hardware synchronization and thereby greatly reducing the hardware requirements of a QKD system. 
			Nonetheless, classical qubit-based synchronization exhibits poor performance in continuous and high-loss systems, hindering its wide applicability in various scenarios. Here, we propose a qubit-based distributed frame synchronization method that can achieve time recovery in a continuously running system and resist higher losses. Experimental results show that the proposed method outperforms the advanced qubit-based synchronization method Qubit4Sync in a continuously running system. Particularly, the results demonstrate that our method surpasses all previous works in key parameters, including frequency and the synchronization length. 
			We believe our method is applicable to a broad range of QKD scenarios, including drone-based QKD and quantum network construction.

	\end{abstract}
	
	\maketitle
	
	\section{Introduction}

Quantum key distribution (QKD) is a secure protocol that enables two communicating parties to share a secret key over an insecure channel. Since the proposal of the first QKD protocol in 1984 \cite{1984.BENNETT.C.H.Quantum}, QKD has achieved significant advancements in   fiber links  \cite{2010Chentengyun,2012wangshuang,2018Yuan,2018Boaron2,2021zhou-MDI,2021.D.Ma.Opt.Lett,2022Tang-RFI,2022.Lufengyu.Optical,2022.Wangshuang.NaturePhotonics,2023.Lufengyu.Optica,2023.Lo.Hoi-Kwong.Phys.Rev.Lett,2023.Lufengyu.PhysRevLett} \blk, free-space links \cite{2017Takenaka-free,2017Liao2,2020chenhuan,2020.Hua-Ying.Liu.National.Science.Review,2023.BassoBasset.QuantumScienceAndTechnology,2023.Andrej.npjQI}, and network structures \cite{2019Dynes-network,2021Chenyuao-network,2022.Fan-Yuan:Optica,2014.WangShuang:Opt.Express,2011.M.Sasaki:Opt.Express}. Remarkably, long-distance communication over a 1000 km fiber spool \cite{2023.LiuYang.PhysRevLett} was achieved using   twin-field QKD  \cite{2018Lucamarini,2018Wang-TF} \blk.

Clock synchronization is a crucial step in the normal operation of QKD systems. Existing QKD demonstrations generally synchronize their clocks using the global navigation satellite system \cite{2007.Schmitt-Manderbach.PhysRevLett.98.010504,2015.Vallone.PhysRevA.91.042320,2021.Avesani.npjQuantumInformation}, electrical cables \cite{2013Wei-QC,2017Islam-qudits,2022Ioannou2022receiverdevice,2020Wei,2022.GU:Science.Bulletin}, or classical light \cite{2022Avesani-light,2023.LiWei:NaPho,2023Gru-highrate}. All of these methods require additional hardware, increasing the complexity of the QKD system. 

In 2020, Calderaro et al.~\cite{2020.Luca.Calderaro.PRApplied} introduced the Qubit4Sync method, wherein Alice transmits an initial public synchronization string in the first state, and time synchronization is accomplished by Bob, by post-processing the detection events. Consequently, synchronization can be achieved by transmitting quantum states between Alice and Bob, thus eliminating the necessity for hardware synchronization. This method has been demonstrated in realistic QKD systems \cite{2020.Costantino.Agnesi.Optica,2021Avesani}, and several variants have been   reported~\cite{2021-Wang-qubit,2021.Cochran.Roderick.D.Entropy,2022.Wang.Tao.PhysRevApplied,2021.Wang.OptExpress}. \blk

However, Qubit4Sync has limitations that hinder its wide applicability. First, Qubit4Sync encodes a synchronization string at the beginning of the quantum states. The inherent jitter of the system clock may cause the clock recovery process to fail. When clock synchronization fails, Alice must reestablish the qubit-based synchronization process and becomes unable to distribute key bits with Bob continuously. Second, to compensate for the channel loss, the length of the synchronization string must increase with higher losses to achieve successful synchronization. For example, at a channel loss of approximately 15 dB, a synchronization string of length $10^6$ is required, whereas at channel losses exceeding 34 dB, the length is increased to $10^7$~\cite{2020.Costantino.Agnesi.Optica}.
	
	\begin{figure*}
		\centering
		\includegraphics[width=0.9\linewidth]{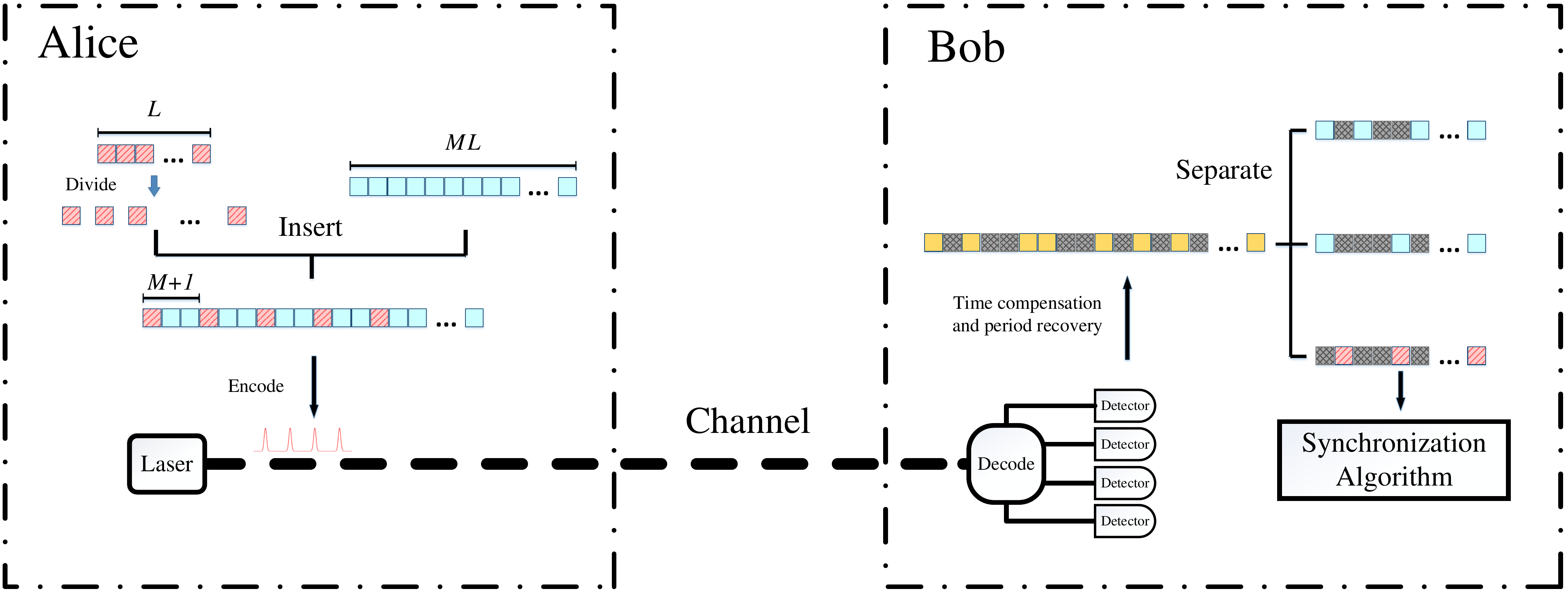}
		\caption{Schematic of QKD based on qubit-based distributed frame synchronization. Alice divides the synchronization string into individual bits and inserts them into random strings to form synchronization frames, and uses the synchronization frames to encode light pulses. The red and blue squares represent synchronization and random bits, respectively. After the encoded light pulse is transmitted through the channel, it is first decoded by Bob into four quantum states and record time of detections. Bob recovers the period $\tau^B$ of the pulse through these detections, and separates the consecutive detections according to $\tau^B$ to form Bob's receive string. We use yellow squares to represent bits of the received string because Bob cannot confirm the source of the received bits. Bob joins the portions of the received string according to the structure of the synchronization frame constructed by Alice. The separated detections encoded by the synchronization string are used for clock synchronization, and the random string detections are required for the QKD protocol.}
		\label{fig_AliceBob_Synchronization}
	\end{figure*}

	Inspired by the distributed frame synchronization method used in classical communication \cite{2000.Adriaan.J.IEEE}, we propose a qubit-based distributed frame synchronization method that can overcome the above limitations in Qubit4Sync. In our method, Alice periodically inserts a segment within the synchronization
	string into quantum states. This allows Bob to execute the qubit-based synchronization algorithm and distill the secure key bits continuously and simultaneously, even in the presence of the inherent jitter of the system clock.

	  The proposed qubit-based distributed frame synchronization method was tested in a polarization-based BB84 system. Experimental results demonstrate that the system, operating at a repetition rate of $625$ MHz can establish synchronization using the proposed method with a synchronization string of length $L=10^4$ under a maximum transmittance loss of 29.2~dB. \blk  We compared our distributed frame synchronization method with Qubit4Sync in the same system, operating in continuously running mode. Experimental results show that when the system operation at a frequency of 50 MHz and time exceeds 80 s, our method maintains a low quantum bit error rate and hence extracts secure key bits, whereas Qubit4Sync fails under these conditions.   Furthermore, our method exhibits stability over a one-hour stability test. \blk In addition, we demonstrate that our method can execute time offset recovery successfully without lengthening $L$ at a larger channel loss via post-processing.

	The remainder of this paper is organized as follows: In Sections~\ref{Robust and fast qubit-based frame synchronization algorithm} and \ref{Frame synchronization under high loss}, we describe our synchronization algorithm in detail. Section~\ref{EXPERIMENT AND RESULTS} presents our experimental results. Finally, we summarize our work in Section~\ref{Conclusion}.

	\section{Description of our qubit-based synchronization method}\label{Robust and fast qubit-based frame synchronization algorithm}

Fig.~\ref{fig_AliceBob_Synchronization} shows the schematic of a QKD system based on our proposed qubit-based distributed frame synchronization method. In conventional QKD systems  Alice only encodes random bits into the quantum states for distilling secure key bits and the time synchronization requires additional hardware. Conversely, in our setup, Alice must encode the quantum states with a public synchronization string for synchronization and random bits for key bit distillation. The time between two consecutive qubits $\tau^A$ is set by Alice's clock. The encoded quantum states are then sent to Bob over a lossy channel.

 Owing to channel loss, Bob receives some of the quantum states and analyzes the received qubits, measuring the arrival time using his clock. 
 Bob's main goal is to determine the positions of the detected qubits in Alice's random bits. This crucial step is essential for correctly generating the raw key bits, and subsequently, the sifted key. To do so, Bob must accomplish three tasks: $(i)$ Recover   Bob's period $\tau^B$ \blk  from the detections. $(ii)$ Separate the detection of the synchronization string from the random string. $(iii)$ Calculate the time delay between the sent and measured strings. Below, we describe how these three tasks can be accomplished using only the qubits exchanged during the QKD protocol, without requiring additional hardware.

	
	\subsection{Generation of qubits for synchronization and key distribution}\label{The construction of synchronization frame}

	We begin by describing how Alice generates a frame for synchronization and key distribution. As shown in Fig.~\ref{fig_AliceBob_Synchronization}, Alice generates a synchronization string $s^A$ (the red string) with a total length $L$ and a random bit string (the blue string) with a length of $M \times L$ where $M$ represents the ratio of random bits to the number of synchronization bits in one synchronization frame. $s^A$ is public and known to Bob; however, the random bit string is used to generate secure key bits, which are kept secret from Bob. Alice disperses the synchronization string of length $L$ into individual bits and evenly inserts them into the random string to form a frame of length $N_f=(M+1)L$. If Alice must produce several frames, she reuses $s_A$ and refreshes the random string to meet security demands.

	Similar to the method used in Qubit4Sync \cite{2020.Luca.Calderaro.PRApplied}, $s^A$ is a particular string in which each bit $s_i^A$ has a value of $+1$ or $-1$ and the total string has an autocorrelation $x^{AA}_m$ satisfying
	\begin{equation}
		\begin{aligned}
			& x^{AA}_0=1, \\
			& x^{AA}_{jL_1} \simeq c_0, ~for~ j > 0, \\
			& x^{AA}_{u+jL_1} \simeq 0, ~for~ u > 0,
		\end{aligned}
	\end{equation}
	where $1>c_0>0$, $u=0,1,...,L_1-1$ and $j=0,1,...,N_1-1$. $N_1$ is the number of periodic peaks and $L_1 N_1 = L$.   The autocorrelation $x^{AA}_m$ is defined as 
	\begin{equation}
		x^{AA}_m= \frac{1}{L} \sum_{n=0}^{L-1}s^{*A}_{n+m}s^A_n,
	\end{equation}
	where $m=0,1,...,L-1$. \blk  $s^A$ can be generated using the following equation: 
	\begin{equation}
		s^A_{u+jL_1}=2 \Theta (y_{u,j}-\lambda x_u)-1,
	\end{equation}
	where $\Theta (x)$ is the Heaviside function and $x_u$ and $y_{u,j}$ are random numbers uniformly distributed between 0 and 1. The parameter $\lambda$ is used to tune the value of $c_0$. If $\lambda \leq 1$, we have $c_0 = \lambda^2/3 $, whereas if $\lambda > 1$, $c_0 = 1-2\lambda/3$.

	Alice encodes her quantum states based on the previously generated strings. If the individual bits are part of $s^A$, Alice assigns the values $+1$ and $-1$ to H and V polarizations in $Z$ bases. If the bits are part of the random string, Alice encodes the four polarizations based on the BB84 protocol. Finally, the encoded quantum states are sent to Bob via a lossy quantum channel.

	Upon arrival at Bob's location, the qubits are analyzed using a decoder, and their arrival time is measured using Bob's clock. Although Bob receives some of the quantum states, channel loss and detector dark counts prevent him from determining whether the detected signals originate from the synchronization string $s^A$ or a random string. Initially, Bob cannot accurately determine which detected qubits belong to Alice's random bits. Next, we describe in detail how Bob can accomplish this by establishing clock synchronization based solely on the detections, making use of the autocorrelation properties of the synchronization string. That is, Bob must first recover the period $\tau^B$ from the detections, then separate the detection of the synchronization string from the random string, and finally calculate the time delay between the sent and measured strings using the autocorrelation properties of the synchronization string.
	
	\subsection{Period recovery and time compensation}\label{Period recovery}

	In this subsection, we describe the process of recovering the period $\tau^B$ and compensating for the constant time delay caused by detectors in the detections. It is important to note that Bob's period recovery plays a crucial role in clock synchronization between Alice and Bob. Knowing the exact value of Bob's period $\tau^B$ enables him to accurately reconstruct the separations in the raw key between consecutive detections.

	To describe the synchronization algorithm clearly, we define $t^m_a$ as the measured arrival time of qubits according to Bob's clock, where $a \geq 1$ enumerates the obtained detections by Bob. Bob predicts $t^e_a$ as the expected arrival time, expressed as 
	\begin{equation} t^e_a=t_0+n_a\tau^B+\epsilon_{a}, \quad n_{a} \in {\mathbb{N}}, 
	\end{equation} 
	where $t_0$ is the initial time offset and $n_a$ denotes the position of the sent qubit in Alice’s raw key. $\epsilon_{a}$ is a normal random variable with variance $\sigma^2$ and zero mean. Bob's goal is to obtain $t_0$ and $\tau^B$, which can be calculated from Bob's detections and the exposed synchronization string.

	We first describe how Bob recovers the period $\tau^B$ from the detections. Following an approach similar to that of  Qubit4Sync~\cite{2020.Luca.Calderaro.PRApplied}, Bob samples the arrival times of photons and applies a fast Fourier transform to the sampled data to obtain an initial estimate $\tau^B_0$ for $\tau^B$. The sampling rate is $4/\tau^{A}$, where $\tau^A$ represents the time at which Alice sends the light pulse. To expedite computation, the number of samples is limited to $N_s=10^6$, resulting in an error between $\tau^B_0$ and Bob's actual period $\tau^B$ of approximately $4/\tau^{A}N_s$. In general, $\tau^B_0$ cannot satisfy the accuracy requirements for period calculation. To obtain a more accurate value, Bob applies a least-trimmed-squares algorithm of   mod${\tau_0^B}(t_{a+b}^m)$ \blk as a function of $t_{a+b}^m$ and obtains the slope $k$ of the linear model. The more accurate period for Bob, denoted by $\tau^B$, is given by $\tau^B=\tau^B_0/(1-k)$.

	The precisely calculated $\tau^B$ allows us to predict the measured time $t^m_a$ more accurately, which satisfies
	\begin{equation}\label{Eq_Gate}
		\begin{aligned}
		\frac{1}{D} \sum_{b=1}^{D} |\mathcal{E}^I_a (b)|^2 \simeq \sigma_g^{2} .
		\end{aligned}
	\end{equation}
	where $D \in \mathbb{N}$ is the number of pulses detected by Bob, $\mathcal{E}^I_a (b)$ is the time-interval error between two different detections $a$ and $a+b$ and $\mathcal{E}^I_a (b) = (t^m_{a+b} - t^e_{a+b}) - (t^m_a - t^e_a)$. $\sigma_g$ is the gate width for post-processing, and all detection events outside the gate are discarded.

	When Bob obtains an accurate estimation of $\tau^B$, he performs post-processing to compensate for the time deviation caused by photons arriving at different detectors via various paths. We remark that this step is crucial for implementing qubit-based synchronization in real-life systems, but has not received sufficient consideration in previous studies. As shown in Fig.~\ref{fig_AliceBob_Synchronization}, in a typical QKD system, the photons are detected by four detectors. Owing to inherent electronics or path differences, the click times of the four detectors are not exactly the same, even when using the same trigger. This discrepancy can result in incorrect clock synchronization. If the absolute value of the time deviation between two detectors is greater than $\sigma_g$, they cannot satisfy Eq.~\ref{Eq_Gate} simultaneously. If it exceeds $\tau^B/2$, the two detectors cannot recover the time offset simultaneously.

	To compensate for electronics and path differences in the system, we can calculate the time deviations of the four detectors based on Bob's detections. We denote the time of detection events corresponding to the four polarizations as $t_a^{m,P}$ for $P\in{H,V,D,A}$. We then perform linear regression of mod $_{\tau^B}(t_a^{m,P})$. The intercept of the linear model for each polarization on the vertical axis corresponds to the time deviations introduced by the different paths. We use $t_d^P$ to represent the time deviations, and define $t_d^H$ as 0. The time deviations $t_d^P$ for the other polarizations can be calculated using the difference in the intercept between the corresponding linear model and the linear model of polarization $H$. We can compensate for the time deviations in $t^m_a$ directly through post-processing.

	\subsection{Synchronization string separation and time offset recovery}\label{Synchronization string distinguish}

	After successfully recovering the period $\tau^B$, Bob can correctly separate consecutive detections and assign the values $+1$ and $-1$ to the two orthogonal $Z$-basis states. In the event of $X$ bases or no detection, he assigns the value 0. Finally, Bob can produce a string $s^B$ of length $N_f$ with values 0, $-1$, or $+1$. 
	
	However, owing to channel loss and detector dark counts, Bob cannot immediately determine which detections come from the synchronization string and which are from the random string. This makes it impossible for Bob to calculate the time offset by performing a cross-correlation operation directly. Therefore, to calculate the time offset, Bob must first determine which detections come from the synchronization string and reconstruct the synchronization string $s^B_{syn}$ from these detections, and then calculate the time offset using the reconstructed synchronization string $s^B_{syn}$.

\begin{figure}
		\centering
		\includegraphics[width=\linewidth]{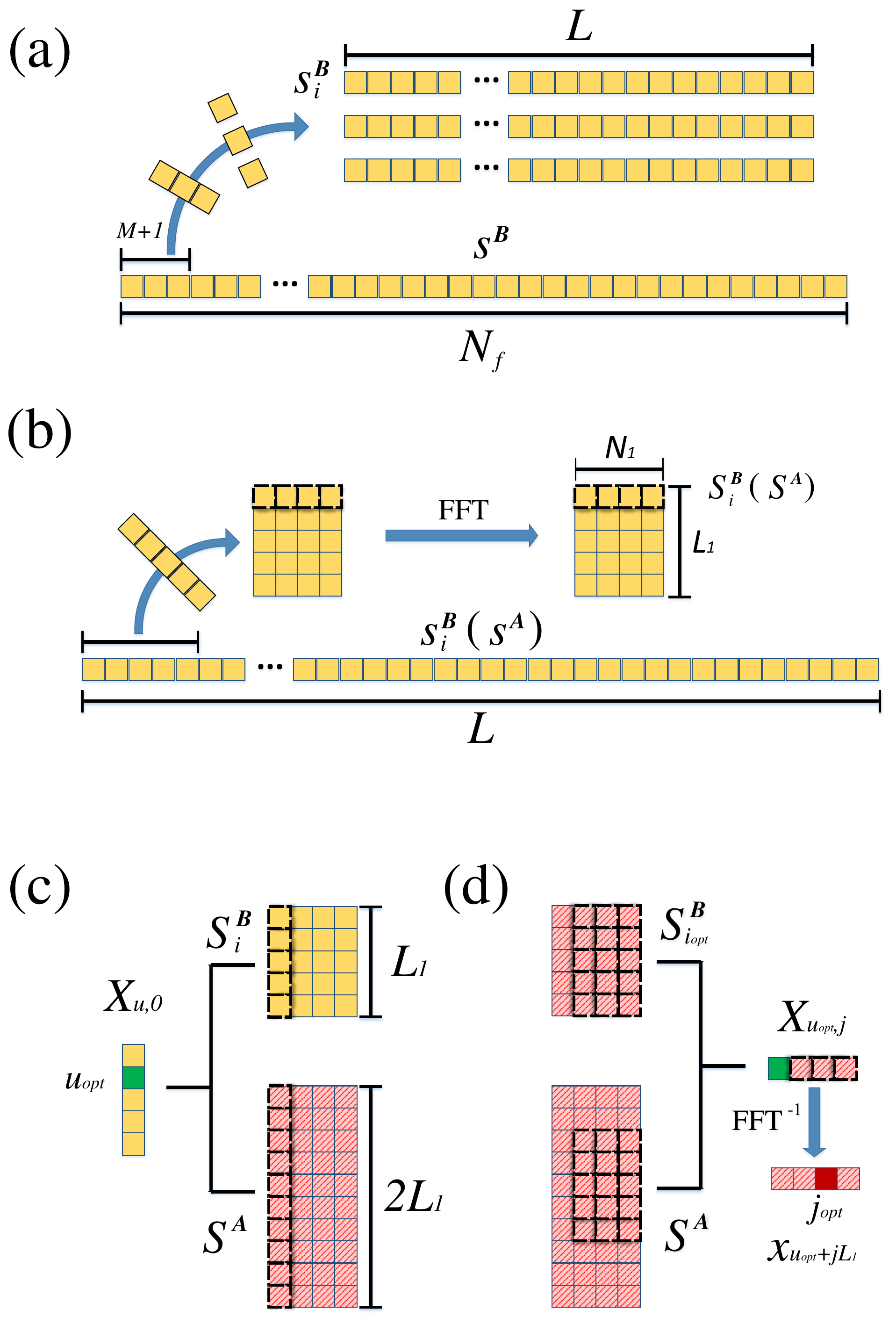}
		\caption{Schematic of synchronization string separation and time offset recovery. Yellow squares represent undetermined bits, and red squares represent the synchronization string and detections from the synchronization string. (a) Bob selects a received string $s^B$ of length $N_f$, and extracts $M+1$ separated strings $s^B_i$ from it according to the structure of the synchronization frame. (b) Bob rearranges the synchronization string $s^A$ and all $s^B_i$ into a two-dimensional matrix of order $L_1 \times N_1$ and applies a fast Fourier transform to each row, denoted as $S^A$ and $S^B_i$. (c) Cross correlation is made between the first column of all $S^B_i$ and the first column of $S_A$, and $i_{opt}$ and $u_{opt}$ corresponding to the maximum cross correlation value are recorded. (d) According to $i_{opt}$ and $u_{opt}$, the remaining ($N_1$-1) columns of $S^B_{i_{opt}}$ and $S_A$ are multiplied and combined with $X_{u_{opt},0}$ obtained from step (b) to form an array $X_{u_{opt},j}$. The row cross correlation $X_{u_{opt}+jL_1}$ is obtained by inverse fast Fourier transform (FFT$^{-1}$), and the corresponding $j_{opt}$ when the cross-correlation value was maximum is recorded.}
		\label{fig_processTO}
	\end{figure}

	We first describe how Bob identifies the detections of the synchronization string. As shown in Fig.~\ref{fig_processTO} (a), Bob first separates strings of $M+1$ adjacent bits into blocks, producing $L$ blocks of length $M+1$ which are
	fitted into an $(M+1)\times L$ matrix. Each row of the	matrix constructs a string $s^B_i$ of length $L$, where $i\in\{1,2,3...(M+1)\}$. Here, we use yellow squares to represent the elements of $s^B$, because Bob cannot confirm whether the detections are from the synchronization string or random bit string. Based on the method described in Sections~\ref{The construction of synchronization frame}, one row of the constructed matrix has a maximal correlation with $s^A$; however, Bob does not know which row that is. Bob's next goal is to determine the optimal row and use it to calculate the time offset. 
	
	Next we describe how Bob finds the optimal row, which has index $i_{opt}$ and contains the string $s_{syn}$. Bob first rearranges the synchronization string as shown in Fig.~\ref{fig_processTO} (b), and applies a fast Fourier transform to each row of all the rearranged matrices. The number of rows in the synchronization matrix $S^A$ is doubled to reserve space for the lookup of $i_{opt}$ and offset calculations. 
	
	To calculate $i_{opt}$ quickly, Bob cross-correlates the first column of $S^A$ and all $S^B_i$ and denotes the result as $X_{u,0}$, as shown in Fig.~\ref{fig_processTO} (c). The matrix $S^B_i$ corresponding to the maximum cross-correlation peak value $X_{u_{opt},0}$ is $S^B_{i_{opt}}$. Bob records $i_{opt}$ and $u_{opt}$ for calculation of the offset. Finally, Bob conducts a horizontal cross-correlation operation between $S^B_{i_{opt}}$ and $S_A$ according to the optimal row $u_{opt}$, and the position corresponding to the peak of the cross-correlation is the optimal column $j_{opt}$. 
	
	The time offset can be calculated as $\tau^B[i_{opt} + u_{opt}(M+1) + j_{opt} L_1 (M+1)]$, based on the construction coefficient $M$ of the synchronization frame, the period $\tau^B$ and the position $i_{opt}$, $u_{opt}$, $j_{opt}$ corresponding to the optimal value of the cross-correlation.

	\subsection{Computational complexity}\label{Efficient frame synchronization}

In this subsection, we analyze the computational complexity of the proposed frame synchronization algorithm. After the synchronization string separation and time offset calculation steps, as illustrated in Fig.~\ref{fig_processTO}, we can estimate the computational complexity step by step as follows.

In steps (a) and (b), Bob rearranges $s^B$ into $M+1$ matrices and performs FFT operations on all $s^B_i$. Each $s^B_i$ consists of $L_1 \times N_1$ elements, so the total computational cost is $\mathcal{O}[(M+1) L_1 N_1 \log_2 N_1]$.

In step (c), Bob needs to determine the positions $i_{opt}$ of synchronization string detections and $u_{opt}$ of optimal column cross-correlation, which requires calculating the cross-correlation between the first column of all $S^B_i$ and the first column of $S_A$. The calculation of cross-correlation in step (c) can be simplified using FFT, with a computational cost of $\mathcal{O}[(M+1)L_1 \log_2 L_1]$.

In step (d), Bob needs to calculate the optimal row cross-correlation position $j_{opt}$ of $S^B_{i_{opt}}$ and $S_A$, with a computational cost of $\mathcal{O}(N_1 \log_2 N_1)$.

Hence, with $N_1 = \log_2 L$ and $L_1 N_1 = L$, the total computational cost for Bob is 	
\begin{equation}\label{eq_RateCD}
	\mathcal{O}[(M+1)Llog_2 (log_2 L)].
\end{equation}

Compared to Qubit4Syn, which has a computational cost of $\mathcal{O}[L\log_2 (\log_2 L)]$, the disadvantage arises from steps (a) and (b), where our method requires searching for the location of synchronization string detections.
	 
	\blk
	
	\section{Distributed frame synchronization resisting high loss}\label{Frame synchronization under high loss}
	From Section~\ref{Robust and fast qubit-based frame synchronization algorithm}, we note that calculating the time offset is a key step that requires  sufficient detections by Bob to ensure that the maximum correlation peak is distinguished from the others. When the overall transmittance loss is excessively large, the time offset calculation may fail.   To enhance the probability of successful execution of offset recovery, the value of $L$ should increase as the loss increases, which sacrifices the final secure bit rate. Calderaro et al. showed that the minimum synchronization string length $L$ satisfies $L \simeq 100/\eta$ in ideal conditions with no errors or dark counts~\cite{2020.Luca.Calderaro.PRApplied}, where $\eta$ is the overall transmittance. For instance, given a synchronization string of length $L = 10^6$, Qubit4Sync can cope with a total loss of 40 dB. \blk 

Unlike Qubit4Sync, our distributed frame synchronization method generates each frame using an identical synchronization string, although the synchronization string $s_{syn_i}^B$ reconstructed from $K$ frames would differ owing to the probabilistic loss of photons in a lossy channel, where the subscript $i\in\{1,2,...K\}$. Because they all originate from $s^A$, Bob can accumulate all the information from  $s_{syn_i}^B$, forming a new string $s_{all}^B$,  to resist channel loss without increasing the length of the synchronization string. 

The detailed method is explained as follows. First, Bob extracts $K$ frames of length $N_f$ continuously from his detections, which all contain complete synchronization frame detections. Because these strings are contiguous, the synchronization string is in the same position in all synchronization frames. This allows Bob to supplement missing detections with detections of the same location in other synchronization frames. In the case of multiple detections occurring at the same location, the results with the highest frequency are used. The time offset  between Bob and Alice is estimated by calculating the correlation between the reconstructed strings.

	Our method shares some steps with Qubit4Sync, but it distinguishes itself by collecting data from $K$ synchronization frames to resist losses. Consequently, Bob accumulates a synchronization string with a total length of $KL$, where the minimum $L$ satisfies $L \simeq 100/K\eta$. Thus, our method efficiently resists high losses.
	
	\blk

	\section{Experiment and results}\label{EXPERIMENT AND RESULTS}

\begin{figure}
	\centering
	\includegraphics[width=\linewidth]{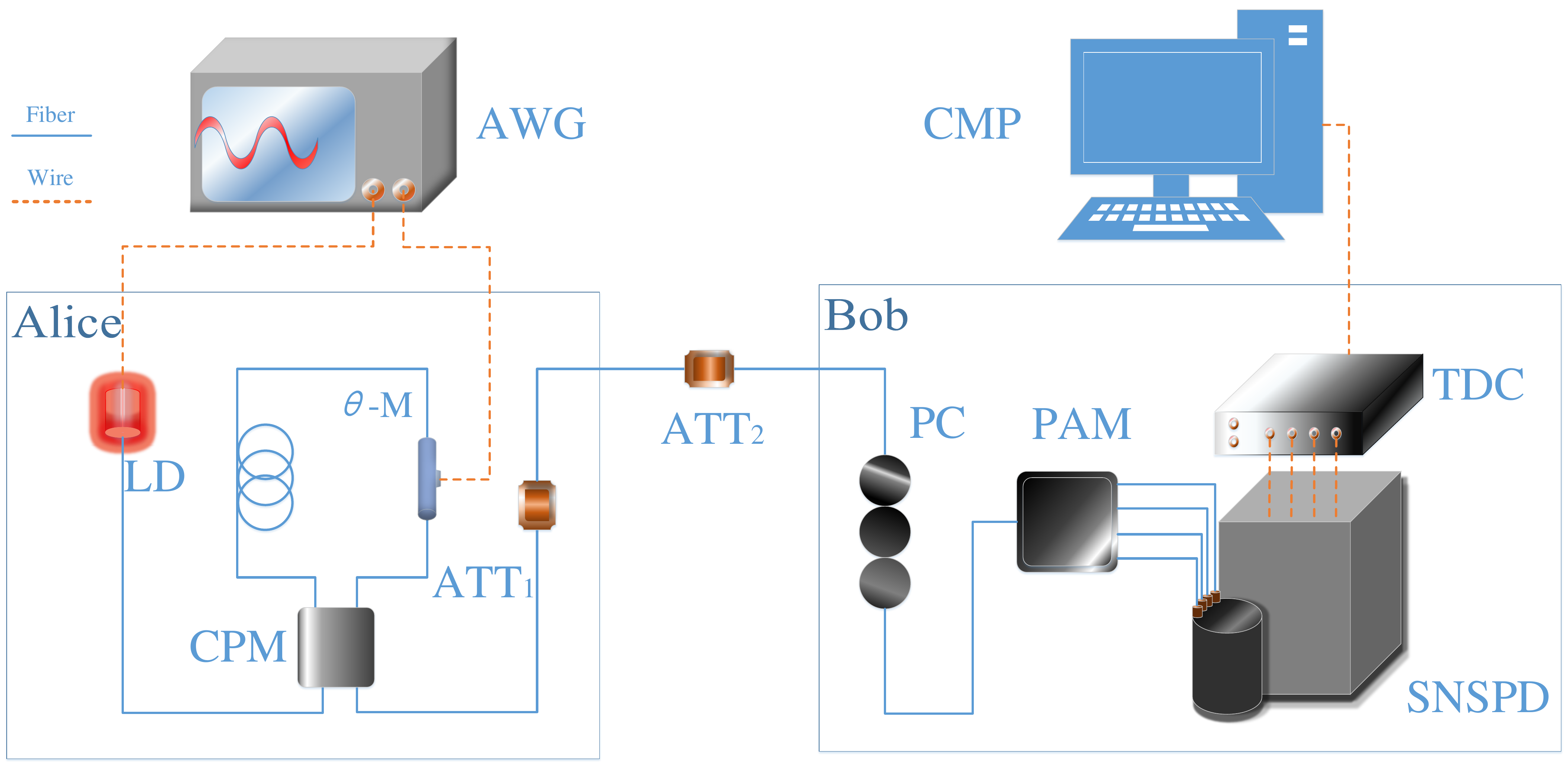}
	\caption{ Experimental setup. AWG: arbitrary waveform generator;  LD:  commercial laser diode; $\theta$-M: phase modulator; CPM: customized polarization module; ATT: optical attenuator; PC: polarization controller; SNSPD: superconducting-nanowire single-photon detector; PAM: polarization analysis module; TDC: time-to-digital converter; CMP: a computer for synchronization algorithms.}
	\label{fig_System}
\end{figure}

	We tested our qubit-based distributed frame synchronization method by modifying a polarization-based QKD system \cite{2022Chen,2022Huang}. The experimental setup is illustrated in Fig.~\ref{fig_System}.   Alice used commercial lasers (LD, WT-LD200-DL, Qasky Co. LTD) to generate light pulses. \blk Alice's time reference was provided by an arbitrary waveform generator (AWG), and the polarization was encoded through a Sagnac-based polarization modulator (Sagnac PM). ATT$_1$ attenuated the encoded quantum light to the single-photon level, whereas ATT$_2$ was used to simulate channel loss.	
	
At the receiving station, a polarization controller (PC) was used to compensate actively for polarization drift during transmission over the channel. The received quantum states were decoded by the polarization analysis module (PAM) and detected by superconducting-nanowire single-photon detectors (SNSPD) with an average efficiency of $51.3\%$ and a dark count rate of $2 \times 10^{-6}$. The receiver measured the polarization on the Z base with a probability of $90\%$. The detection events were recorded using a time-to-digital converter (TDC) triggered by an internal clock. Subsequently, the detection events were processed using synchronization algorithms implemented on a computer (CMP).

		\begin{table}[]
		\centering
		\caption{Synchronization results under different losses   at different system clock\blk. Loss is the total loss between Alice and Bob, and QBER is the quantum bit error rate. The noise filter gate width $\sigma_g = \sigma$.}
		\begin{tabular}{@{} cccccc @{}}
			\hline \hline
			~~ &Frequency & ~Loss(dB)~ & ~~QBER~~~ & ~~~~$\tau^B (ns)$~~~~ &~~
			\\ \hline
			&\multirow{ 5}{*}{50 MHz} & 17.6 & 0.459\% &20.0000184 & 
			\\ 
			&                         & 20.0 & 0.617\% & 20.0000182 &
			\\ 
			&                         & 22.7 & 0.457\% & 20.0000186 &
			\\ 
			&                         & 26.5 & 0.788\% & 20.0000185 &
			\\ 
			&                         & 29.7 & 0.625\% & 20.0000185 &
			\\ \cline{2-5}
			&\multirow{ 5}{*}{625 MHz} & 23.5 & 2.37\% & 1.60000742 &
			\\
			&                          & 25.5 & 2.17\% & 1.60000747 &
			\\
			&                          & 27.1 & 2.03\% & 1.60000753 &
			\\
			&                          & 28.4 & 2.11\% & 1.60000744 &
			\\
			&                          & 29.2 & 1.88\% & 1.60000749 &
			\\
			\hline \hline
		\end{tabular}
		\label{table_RCR_QBER}
	\end{table}
	
\begin{figure}
		\centering
		\includegraphics[width=0.8\linewidth]{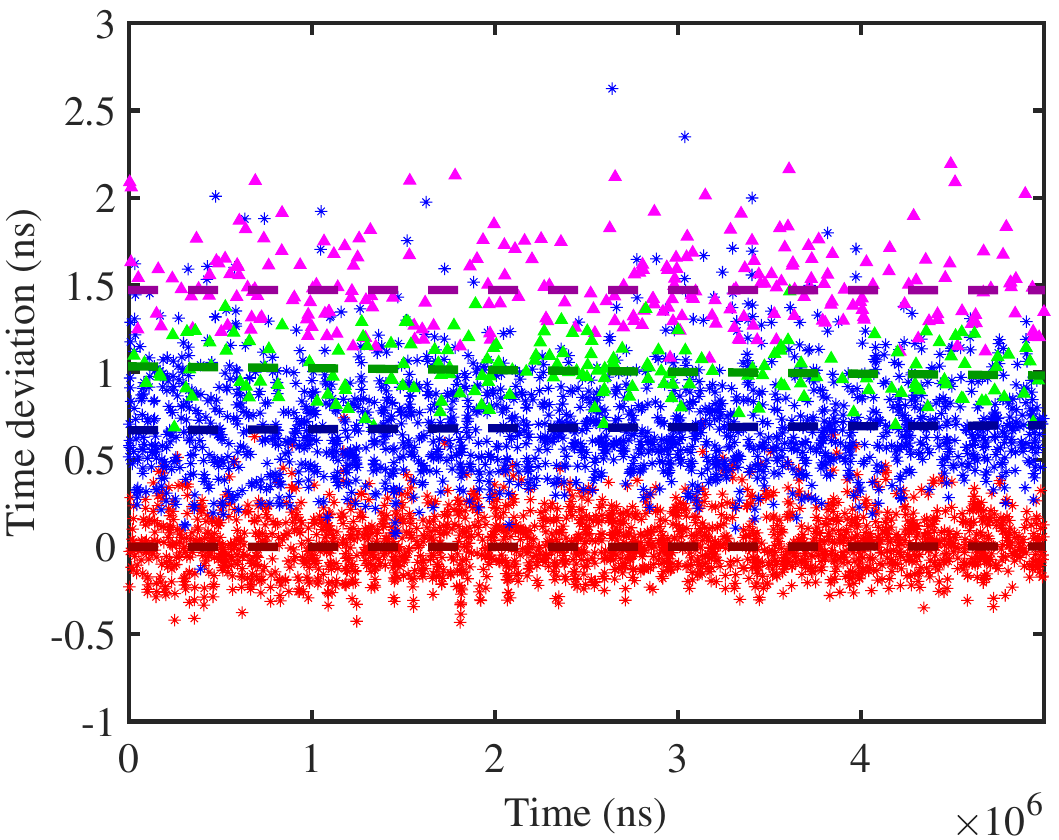}
		\caption{Distribution of time deviation for different polarization detections. The estimated time deviation of the four polarizations H, V, D and A are  represented by red asterisks, blue asterisks, magenta triangles and green triangles, respectively. Solid, dashed, dotted, and dash-dot lines represent average time deviation  for   H, V, D, and A\blk, respectively.  }
		\label{fig_LTSPol}
	\end{figure}

We first tested the time compensation method described in in Section~\ref{Period recovery}.   At a repetition rate of 50 MHz, \blk we sent  polarization H/V and performed period recovery by analyzing detections. Bob's detections were then sorted into the corresponding time windows based on the recovery period $\tau^B$. To determine the time deviation introduced by different paths, we calculated mod $_{\tau^B}(t_a^{m,P})$ for the recorded time of four polarizations. Following the method described in Section~\ref{Period recovery}, we can estimated the time deviation caused by the path difference. The results are presented in Fig.~\ref{fig_LTSPol}.  It can be seen that the black solid, dashed, dotted, and dash-dot lines represent obtained average time deviations $t_d^H$, $t_d^V$, $t_d^D$, and $t_d^A$ of 0 ns, 0.65 ns, 1.48 ns, and 1.01 ns, respectively. We  applied obtained time deviations to  post-processing in the following tests.

		\begin{table}[]
		\centering
		\caption{Success probability and QBER in offset recovery test. Loss is the total loss of the system, $P_s$ is the probability of successful execution of offset recovery, QBER is the quantum bit error rate, and $K$ is the number of synchronization frames considered by the complement operation. The length of the synchronization string is $L=10^5$ and the total sampling time at each loss is 20~s.}
		\begin{tabular}{@{} cccccc @{}}
			\hline \hline
			Loss (dB) & $K$ & 1 & 2 & 4 & 8 \\ 
			\hline 
			\multirow{ 2}{*}{26.5} & $P_s$ & 20.3\% & 81.1\% & 98.9\% & 99.4\% \\
			& QBER & 40.0\% & 10.1\% & 1.33\% & 1.08\%
			\\ 
			\hline
			\multirow{ 2}{*}{29.7} & $P_s$  & 5.35\% & 16.4\% & 78.2\% & 98.7\% \\
			& QBER & 47.4\% & 41.9\% & 11.4\% & 1.27\%
			\\ 
			\hline \hline
		\end{tabular}
		\label{table_K_QBER}
	\end{table}

	We then evaluated the frame synchronization performance under various loss conditions and system frequencies using the method described in Section~\ref{Frame synchronization under high loss}.   The results are listed  in Table \ref{table_RCR_QBER}. At a frequency of 50 MHz, \blk we set the average number of photons per pulse sent by Alice to 1 and obtained five different losses by adjusting ATT$_2$. The synchronization process used a synchronization string of length $L=10^5$ and synchronization frame construction factor of $M=1$, with an acquisition time of 100~s. After performing time recovery, we estimated the quantum bit error rate (QBER), with results as presented . Our  method demonstrated consistent recovered time periods and time delays under different losses. This enabled us to obtain raw key bits with a QBER below $0.457\%$, which aligns closely with the intrinsic optical misalignment of the system.

	\begin{table*}[]
		\centering
		\caption{  Comparison of performance of existing qubit-based synchronization method. \blk}
		\begin{tabular}{ccccc }
			\hline \hline
			Reference & Frequency(MHz) & ~~Loss(dB)~~  & ~~Synchronization length~~  &~~QBER~~ 
			\\ \hline
			\multirow{ 2}{*}{Calderaro et al. \cite{2020.Luca.Calderaro.PRApplied}} & \multirow{ 2}{*}{50} & 19.0 & $10^6$ & 0.30\%
			\\ 
			&  & 38.0 & $10^7$ & /
			\\
			\multirow{ 2}{*}{Wei et al. \cite{2023.Wei:Photon.Res}} & \multirow{ 2}{*}{50} & 9.96& $5 \times 10^4$  & 0.653\%
			\\
			&  & 30.0 & $5 \times 10^4$ & 1.36\%
			\\
			Cochran et al. \cite{2021.Cochran.Roderick.D.Entropy} & / & $20^\dagger$  & $4140^\dagger$ & /
			\\
			\multirow{ 2}{*}{Our method} & 50 & 29.7 & $10^5$ & 0.625\%
			\\
			& 625 & 29.2 & $10^4$ & 1.88\%
			\\
			\hline \hline
			$\dagger$ represents a simulated experiment.
		\end{tabular}
		\label{table_Reference}
	\end{table*}

	\begin{figure}
		\centering
		\includegraphics[width=0.8\linewidth]{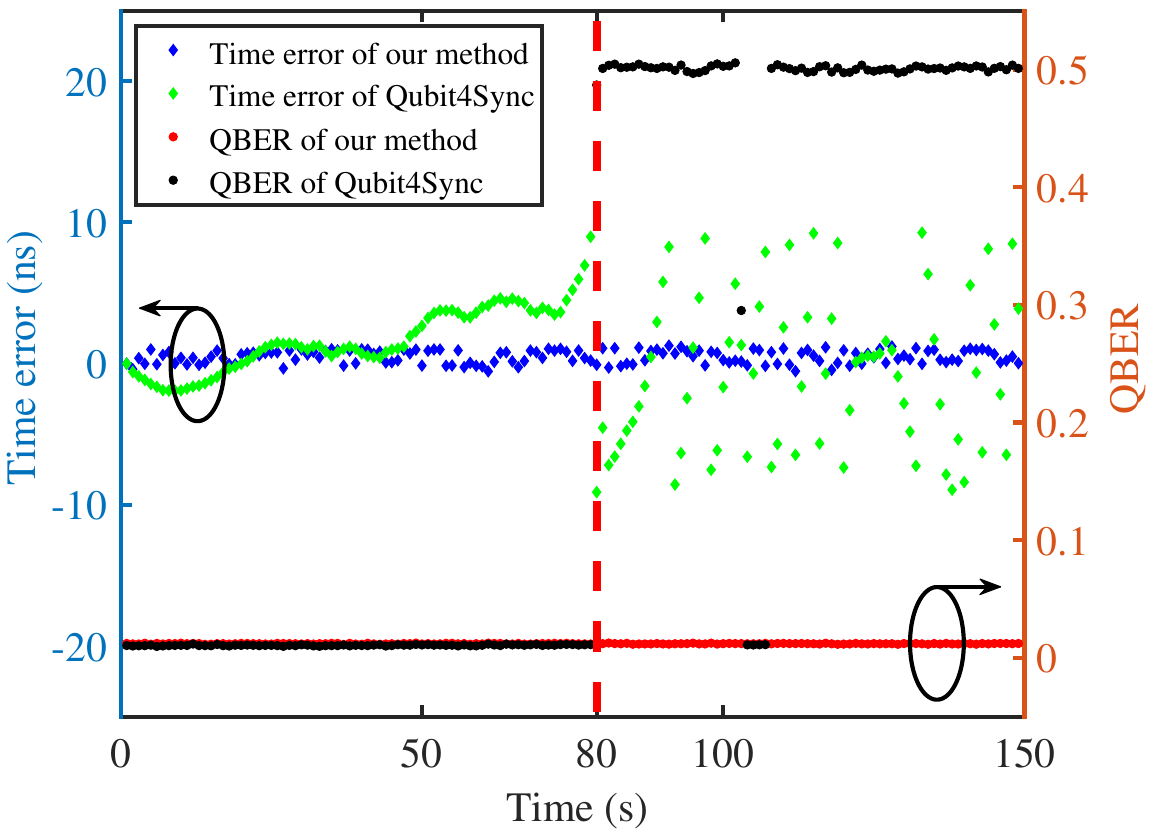}
		\caption{Stability test of Qubit4Sync and our method. The red and black circles represent the QBER of Qubit4Sync and our method, respectively. The blue and green diamonds represent the time  error of Qubit4Sync and our method, respectively. Ideally, the time  error is 0 ns and QBER is 0\%.}
		\label{fig_CIDI_expeimental}
	\end{figure}

	We conducted tests on a continuously running system to demonstrate the stability of our method. For comparison, we also implemented Qubit4Sync using the same system. We plotted the time error,  which defines as mod$\tau^B(\mathcal{E}^I_a (b))$, and QBER versa the continuous acquisition time. The experimental results are shown in Fig.~\ref{fig_CIDI_expeimental}. After the entire system had operated for more than 80~s, Qubit4Sync was unable to recover the clock accurately, resulting in a time error of approximately 10 ns and a QBER of $50\%$.   In contrast, our method continued to operate effectively  for a duration of 150 s.  \blk

\begin{figure}
		\centering
		\includegraphics[width=0.8\linewidth]{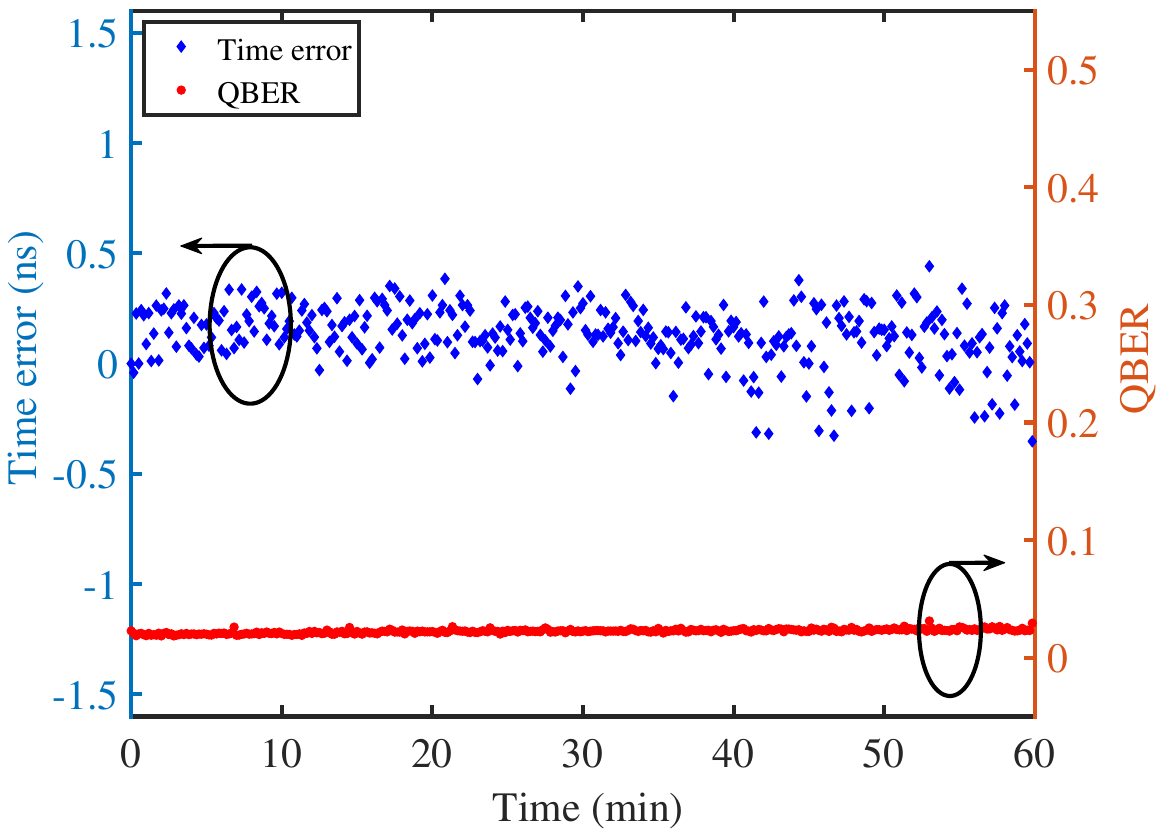}
		\caption{  Stability test of our method at a repetition rate of 625 MHz. The red  circles represent the QBER and the blue diamonds represent the time error, respectively. The  length of synchronization string used for testing is $L = 10^4$ and the total channel loss is 28.4 dB. Each point is calculated by using data collected over a one-second period.\blk}
		\label{fig_625MHz_test}
	\end{figure}

 	We further tested the proposed method in high-speed systems by utilizing an electronic control board  to increase Alice's clock frequency to 625 MHz. We configured the synchronization string length to $L=10^4$ and $M=1$. As indicated in Table \ref{table_RCR_QBER}, our synchronization method exhibits excellent performance even in high-speed systems.  Additionally, to assess the stability of our synchronization method, we maintained a fixed loss of 28.4 dB and conducted a one-hour stability test, the results of which are depicted in Fig.~\ref{fig_625MHz_test}. Over a 1-hour test period, the average QBER was found to be 2.1\% $\pm$ 0.2\%, and the average timing error was $\pm$ 0.37 ns. \blk
	
	\blk
	
	Finally, we tested the synchronization performance improvement of the method described in ~Section~\ref{Frame synchronization under high loss} under the two largest losses described above. We set a constant length of the synchronization string as $L=10^5$ and post-processed the data recorded by Bob with an acquisition time of 20~s. The post-processing method was as described in~Section~\ref{Frame synchronization under high loss}. The results are summarized in Table \ref{table_K_QBER}. At a channel loss of 26.5 dB, with an increasing number of synchronization frames considered by the complement operation described in Section~\ref{Frame synchronization under high loss}, we obtained a lower QBER because of the high probability of the successful execution of time offset recovery. A similar observation applies to the channel loss of 29.7 dB. In particular, at a higher channel loss, our method can execute time offset recovery successfully by accumulating more $K$ frames without needing to lengthen $L$.

	To illustrate the advancements achieved by our results, we compare them with those of the existing Qubit4Sync method, as presented in Table~\ref{table_Reference}. Our method surpasses previous works in key parameters, including frequency and the synchronization length $L$, as evident from the comparison.
	
	\blk

	\section{Conclusion}\label{Conclusion}

	We present a novel qubit-based distributed frame synchronization method inspired by classical communication techniques. Our proposed method addresses the limitations encountered in the Qubit4Sync approach and offers improved performance and robustness. By incorporating periodic quantum state insertions into the synchronization string, we enable the continuous execution of the qubit-based synchronization algorithm and secure key bit distillation, even when confronted with inherent system clock jitter.
	
	The feasibility and robustness of the proposed method were demonstrated through an experimental implementation in a polarization-based BB84 system operating   at a frequency of up to 625 MHz. \blk The experimental results demonstrate the successful establishment of synchronization using our proposed method. We believe our synchronization algorithm is applicable to a broad range of QKD scenarios, including  drone-based QKD~\cite{2022-CLEO-Drone,2023-Tian-Drone}, chip-based systems~\cite{2021Para-chip,2023Du-chip,2023Sax-chip}, and the construction of quantum networks ~\cite{2013.F.Bernd.Dynes.Nature,2023Huang}.
	\section{Acknowledgments}
	This study was supported by the National Natural Science Foundation of China (Nos. 62171144 and 62031024), the Guangxi Science Foundation (No.
2021GXNSFAA220011), the Training plan for Guangxi 1000 Young and Middle-aged Teacher and the Open Fund of IPOC (BUPT) (No. IPOC2021A02). 
	
	\bibliography{DPI_FS}

\end{document}